\documentclass[%
 preprint,
 amsmath,amssymb,
 aps,
]{revtex4-1}

\usepackage{graphicx}
\usepackage{dcolumn}
\usepackage{bm}
\usepackage[flushleft]{threeparttable}
\usepackage{url}
\usepackage{amsmath}
\usepackage{amssymb}
\usepackage{braket}
\usepackage{multirow}
\usepackage{array}
\usepackage{url}
\usepackage{booktabs}
\usepackage{epstopdf}
\usepackage{color,soul}

\newcolumntype{P}[1]{>{\centering\arraybackslash}p{#1}}

\begin{document}

\title{Ultra--fast electric field control of giant electrocaloric effect in ferroelectrics}

\author{Y. Qi,$^1$ S. Liu,$^2$ A. M. Lindenberg,$^{3,4}$ and A. M. Rappe$^1$}
\affiliation{%
 $^1$The Makineni Theoretical Laboratories, Department of Chemistry,\\
 University of Pennsylvania, Philadelphia, PA 19104-6323, United States\\
 $^2$Geophysical Laboratory, Carnegie Institution for Science, Washington, D.C. 20015, United States \\
 $^3$Department of Materials Science and Engineering, Stanford University\\
Stanford, CA 94305, United States\\
 $^4$PULSE institute, 
 SLAC National Accelerator Laboratory, \\
 Menlo Park, CA 94025, United States
 }%
\date{\today}

\begin{abstract}

There is a surge of interest in developing environmentally friendly cooling technology based on the solid--state electrocaloric effect (ECE). 
Here, we point out that negative ECE with a fast cooling rate ($\approx$10$^{11}$ K/s) can be achieved by driving solid crystals to a high--temperature phase with a properly designed electric field pulse. 
Specifically, we predict that an ultrafast electric field pulse can cause a negative ECE up to 35 K in PbTiO$_3$ occurring on few picosecond time scales.
We acquire and analyze these results by clarifying the mechanism of ECE during an adiabatic irreversible process; 
In addition to the conventional explanation of the ECE with entropy reallocation, we simply portray the ECE with the concept of internal energy redistribution.
Electric field does work on a ferroelectric crystal and redistributes its internal energy. 
How the kinetic energy is redistributed determines the temperature change and strongly depends on the electric field temporal profile.
This concept is supported by our all--atom molecular dynamics simulations of PbTiO$_3$ and BaTiO$_3$.
Based on this improved understanding of ECE, we propose strategies for inducing both giant and negative ECE. 
This work offers a more general framework to understand the ECE and highlights the opportunities of electric--field engineering for controlled design of fast and efficient cooling technology.

\end{abstract}

\pacs{Valid PACS appear here}
\maketitle

The electrocaloric effect (ECE) refers to the phenomenon in which the temperature of a material changes under an applied electric field~\cite{Rose12p187604,Takeuchi15p48,Moya14p439}.
Recent years have seen a surge of interest in developing ECE--based cooling technology~\cite{Lu09p1983,Valant12p980,Manosa13p4925,Scott11p229,Neese08p821},
which does not rely on the high global--warming potential (GWP) refrigerants (hydrofluorocarbons and hydrochlorofluorocarbons) that are widely used in traditional vapor compression cooling technology. 
Giant positive ECEs up to 12 K were observed in PbZr$_{0.95}$Ti$_{0.05}$O$_3$, and paving the path toward the practical application of the ECE is a fast--moving research project~\cite{Mischenko06p1270}.
Similar to a mechanical refrigeration cycle, the traditional ECE--based refrigeration cycle involves four steps: electric field application, heat ejection, electric field removal, and heat adsorption~\cite{Akcay07p2909}. 
In order to acquire a higher energy efficiency, the electric field should be removed gradually to avoid irreversibilities~\cite{Plaznik15p57009}, which delays the cooling rate.
Energy efficiency and cooling rate can be greatly improved by taking advantage of negative ECE, in which the temperature of a material decreases upon the application of electric field~\cite{Ponomareva12p167604}. 
In this work, we focus on ultra--fast ECE processes and demonstrate terahertz control of both positive and negative ECEs in ferroelectrics theoretically. 
We first clarify the mechanism of ECE during an irreversible process, 
in which entropy is not conserved but energy is conserved.
The external electric field does work on a ferrolectric crystal and causes structural change, which means a modification in potential energy.
As a result, the kinetic energy, which relates to the temperature, also changes.
The signs and magnitudes of these changes are determined by the electric field profile.
Based on this improved understanding, we demonstrate that negative ECE and giant ECE in prototypical ferroelectrics BaTiO$_3$ and PbTiO$_3$ can be realized with short electric field pulses, 
which correspond to giant cooling rates. 

The ECE has been widely understood as entropy reallocation~\cite{Takeuchi15p48,Valant12p980,Ponomareva12p167604}.
The application of electric field aligns dipoles in a material, 
and the configuration entropy is reduced. 
As a result, the thermal entropy, which corresponds to the lattice vibrations, increases.
However, this mechamism only holds for a reversible adiabatic process, which requires the system to be at equilibrium throughout. 
Ultra--fast ECE, which we will focus on in this study, is typically an irreversible process with non--zero entropy production, 
because the polarization does not have enough time to respond fully to the electric field. 
Besides, for a ferroelectric material system with thermal hysteresis, entropy production is also unavoidable in a loop.
Because of the insufficiencies in explaining ECE in ferrolectrics with entropy reallocation, 
we propose that it is much more straightforward to understand the ECE with the concept of internal energy $U$ (per unit volume) redistribution. 
Here, we should emphasize that from our MD simulation, the volume change is quite small (less than 1\%).
Therefore, the mechanical work is negligible and the internal energy is approximate to the enthalpy.
The work $W$ (per unit volume) done by electric field $E$ is given as 
\begin{equation}
W=\int{\bm{E}\cdot{\rm{d}}\bm{P}}
\end{equation}
where $\bm{P}$ is the macroscopic polarization of the material at finite $T$.
The internal energy change $\Delta{U}$ involves changes in both kinetic energy $\Delta{E_k}$ and potential energy $\Delta{E_k}$
\begin{equation}
W=\Delta{U}=\Delta{E_k}+\Delta{E_p}.
\end{equation}
The temperature change $\Delta{T}$ is associated with $\Delta{E_k}$ as
\begin{equation}
\left<\Delta{E_k}\right>=\frac{3}{2}k_B\Delta{T}N\ \ {\Longrightarrow}\ \ \Delta{T}\propto\left<\Delta{E_k}\right>.
\end{equation}
where $N$ is the number of atoms per unit volume, $k_B$ is the Boltzmann constant, and $\left<\Delta{E_k}\right>$ denotes the ensemble average of the kinetic energy change $\Delta{E_k}$.
Generally, the direction of the polarization change ${\rm{d}}\bm{P}$ is along that of the applied electric field, and therefore $W$ is positive.
In most cases, the applied electric field induces a more polar structure, which possesses lower potential energy for ferroelectric materials ($\Delta{E_p}<0$).   
Therefore, $\Delta{E_k}$ is usually positive when turning on the electric field and negative when removing the electric field, causing heating and cooling respectively. 

However, $\Delta{E_k}$ could be negative upon $E$ field application, thus giving rise to negative ECE.
For example, when there is a phase transition with positive transition energy $E_{\rm{tr}}$ (per unit volume), which is the difference of the potential energies of the two phases,
induced by the applied electric field. $\Delta{E_p}\approx E_{\rm{tr}}$, and if $W<E_{\rm{tr}}$,
we have $\Delta{E_k} = W-E_{\rm{tr}} <0$. In this case, some kinetic energy goes to compensate the transition energy, and the temperature decreases.


Negative ECE is significant because it can offer a fast and direct cooling technique~\cite{Pirc14p17002,Geng15p3165,Ponomareva12p167604}, 
where cooling is achieved through the application of electric field. 
It is one of the rare cases where doing work on a system causes its temperature to decrease. 
We perform MD simulations with bond--valence--model based interatomic potentials, 
which have been proven reliable in simulating structural properties and dynamics of ferroelectrics under various conditions~\cite{Grinberg02p909,Shin07p881,Xu15p79,Liu13p104102},
to illustrate the theory of negative ECE in a realistic context. At 102 K, BaTiO$_3$ crystal is in its rhombohedral phase in our MD simulations. 
An electric field along the (110) direction was applied to drive the system from the rhombohedral to the orthorhombic phase. 
As shown in Fig. 1 (a), the polarization components along the $x$ and $y$ direction ($P_x$ and $P_y$, parallel to $E$) increase a bit (0.04 C/m$^2$),
indicating that the work $W$ done by the electric field $E$ is small. 
The crystal is driven to the orthorhombic phase, 
accompanied by an increase of the potential energy, because there is a transition energy $E_{\rm{tr}}$ for the rhombohedral to orthorhombic phase transition. Since
\begin{equation}
W\approx0,\ \Delta{E_p}=E_{\rm{tr}}>0,
\end{equation}
\begin{equation}
\Delta{E_k}=W-\Delta{E_p}<0.
\end{equation}
we observe in MD simulations a decrease in temperature, with an ultra--fast cooling rate ($\approx$10$^{11}$ K/s). 
Based on the equation (5), which demonstrates that negative ECE requires $W$ to be smaller than $E_{\rm{tr}}$, 
we can estimate the upper limit of electric field $E_{\rm{max}}$ causing temperature decrease in our BTO MD model:
\begin{equation}
W-\Delta{E_p}<0,\  E_{\rm{max}}\cdot\Delta{P}<E_{\rm{tr}}
\end{equation}
\begin{equation}
E_{\rm{max}}<\frac{E_{\rm{tr}}}{\Delta{P}}\approx800\ \rm{kV/cm}
\end{equation}

Similarly, we can also generate a negative ECE through the orthorhombic to tetragonal phase transition.
The case for tetragonal to cubic phase transition is less straightforward, because no unidirectional quasi--static electric field can induce the tetragonal to cubic phase transition.
Simulations demonstrate that the cubic phase of BaTiO$_3$ crystal has a disorder character~\cite{Gaudon15p6,Itoh85p29,Stern04p037601}; 
the dipoles in various unit cells orients in different directions and vary with time, and the macroscopic polarization is zero.
Here, we design a single terahertz pulse~\cite{Qi09p247603,Chen15p6371} which is perpendicular to the tetragonal phase polarization.
A rapidly oscillating electric field pulse can disorder the polarization, effectively changing the system to the cubic phase. 
The post--pulse state behaves as a supercooled cubic phase. We highlight that pulse--induced negative ECE requires the occurrence of a phase transition. 
In the absence of field--driven phase transition, because of the one--to--one relationship between energy and temperature in the same phase, the system with higher energy will have higher temperature.
The temperature decrease due to tetragonal--cubic transition in BTO is larger ($\approx$ 2 K) than that due to orthorhombic--tetragonal transition, 
as shown in Fig.~\ref{f_tc}, which is directly related to the larger transition energy of the tetragonal--cubic phase transition.

From the analysis above, we demonstrate that negative ECE can be achieved with $E$--field--induced phase transitions from the low--temperature phase to the high--temperature phase,
because some of the kinetic energy is lost to compensate the potential energy increase.
Conversely, if the transition is from high--temperature phase to low, temperature increases and we find positive ECE. 

In Fig.~\ref{f_bpto} (a) and (b), we plot the internal energy {\em{vs.}} temperature for BaTiO$_3$ and PbTiO$_3$.
A 600~kV/cm electric field is applied to drive the system from cubic phase to tetragonal phase for each material.
We observe giant ECEs $\Delta{T}=12.7$ K, 58.2 K respectively. 
The temperature change (for the same electric field) for PbTiO$_3$ is about five times of that for BaTiO$_3$. 
This is attributed to its larger transition energy (five times that of BaTiO$_3$) 
and larger polarization change (0.75 C/m$^2$, compared with 0.26 C/m$^2$)~\cite{Al-Saidi10p155304}.

Similarly, we predict large negative ECE in PbTiO$_3$.
An ultra--fast electric--field pulse is applied anti--parallel to the PbTiO$_3$ polarization, as shown in Fig.~\ref{f_bpto} (c).
This electric--field pulse induces some negative local polarization in a positively polarized crystal.
This does work $W=\int{\bm{E}\cdot{\rm{d}}\bm{P}}>0$.
After the pulse, the system passes the energy barrier between the tetragonal and cubic phases, 
and the system evolves to a local minimum corresponding to cubic phase without any external force, as shown in Fig.~\ref{f_bpto} (d).
As expected, this $E$ pulse induces a tetragonal--to--cubic phase transition, and a 35 K temperature decrease is observed, which is much higher than recent experimental observation~\cite{Guyomar06p4491,Sebald07p4021,Lu10p202901,Moya13p1360,Bai10p192902}.
This supercooled cubic phase crystal can be used in a cooling cycle. 
After adsorbing heat and equilibrating with a load, the crystal is contacted with a sink. 
Thermal fluctuation drives it back to its original phase with a higher temperature, and then the crystal gives off heat and cools to its original state.
Such a simplified cycle is triggered by just an $E$ pulse, and is promising in improving the cooling rate and energy efficiency.

It should be emphasized that a triple well free energy landscape, as shown in Fig.~\ref{f_bpto} (d), is necessary for such an $E$ pulse induced negative ECE.
This requires the temperature close to the Curie temperature $T_C$.
Around $T_C$, the free energies of the polar and non--polar states are so close,  
which makes domain wall formation less energetically favorable.
As a result, the applied $E$ pulse tends to rumple the local polarizations and trigger the phase transition, rather than induce domains.

These results indicate that a large transition energy is critical for giant ECE. The transition energy sets the upper limit for negative ECE, according to equations (3) and (5):
\begin{equation}
\Delta{T_{\rm{max}}}=\frac{2E_{\rm{tr}}}{3k_BN}.
\end{equation}
The electric field profile is important for driving the desired phase transition and is determined by the symmetry change accompanied with the phase transition. 
Though PbTiO$_3$ exhibits a giant ECE effect near $T_c$, its high $T_c$ (765~K) may impede practical applications at room temperature~\cite{Shirane51p265}.
Techniques that can suppress $T_c$ of PbTiO$_3$--based ferroelectrics such as doping and strain engineering will be helpful for developing practical negative ECE materials~\cite{Rossetti91p2524}. 



One practical concern is that Joule heating may counteract negative ECE.
Joule heating depends on the conductivity of the sample and the field duration, 
and the conductivity of prototypical ferroelectrics could be affected significantly by soft--mode absorption at THz frequencies.
Therefore, selecting materials with low conductivity is critical for achieving negative ECE with a single pulse. Previous experiments suggest that 
PbTiO$_3$, which we predict to show giant pulse--induced negative ECE in this study, has low conductivity (and heating) under THz excitation.~\cite{Perry64p408}.

In this study, we analyze and explain ECE from an energy point of view:
the electric field does work on a crystal, reallocates its kinetic and potential energies, and causes temperature change.
We propose that negative ECE can be both ultra--fast and giant ($T$ reduction as high as 35 K), and a low--temperature to high--temperature phase transition is required in such an $E$--pulse--induced ECE.
The cooling rate due to pulse-induced negative ECE is fast ($\approx$10$^{11}$~K/s), because of the fast response of polarization in prototypical ferroelectrics to electric field (in picoseconds). 

Y.Q. and A.M.R. were supported by the U.S. Department of Energy, under Grant No. DE--FG02--07ER46431. S.L. is supported by Carnegie Institution for Science.
A.M.L. acknowledges support from the Department of Energy, Basic Energy Sciences, Materials Sciences and Engineering. 
The authors acknowledge computational support from the NERSC of the DOE.

\bibliography{rappecites}
\clearpage
\newpage

\begin{figure}[htbp]
\centering\includegraphics[width=15.0cm]{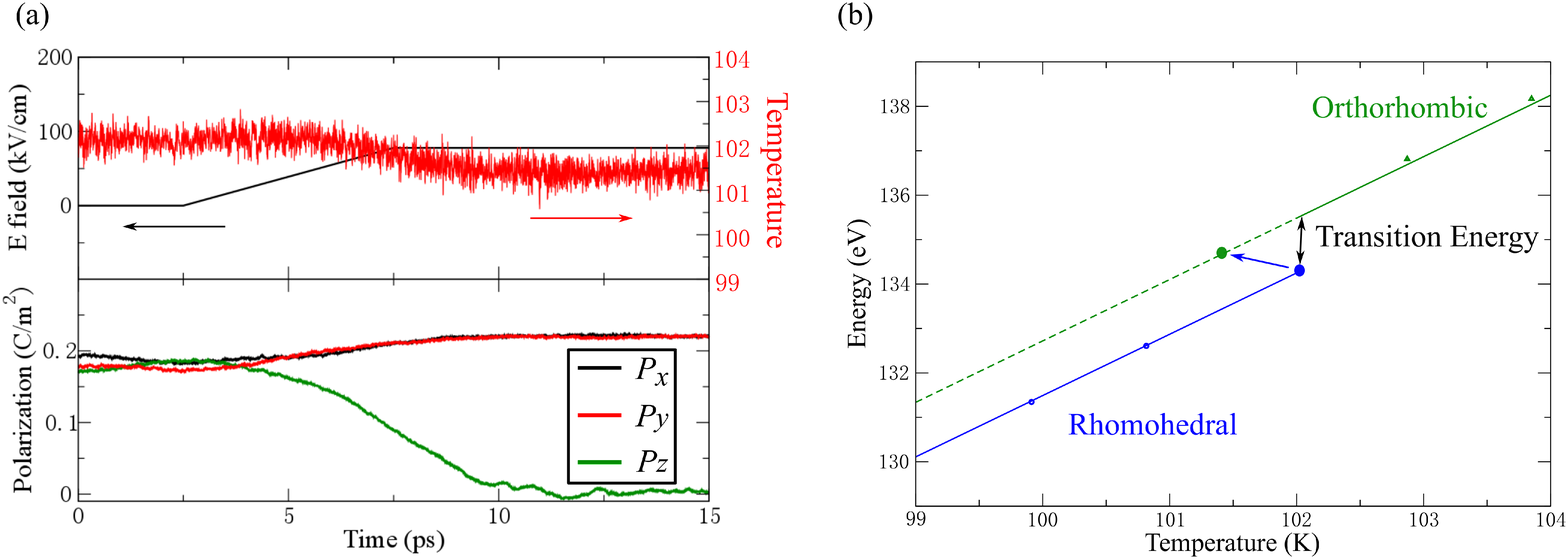} 
\caption{Negative electrocaloric effect associated with the rhombohedral to orthorhombic phase transition in BaTiO$_3$.
(a) The electric field is along the (110) direction. The electric field rises to its steady--state value within 5 ps, rather than instantaneously.
This is because in the negative ECE, less $W$ and entropy production are preferred. 
(b) Schematic plot of internal energy {\em{vs.}} temperature for the two phases, demonstrating the electric field--induced phase transition and ultra--fast temperature reduction.} \label{f_ro}
\end{figure}

\begin{figure}[htbp]
\centering\includegraphics[width=12.0cm]{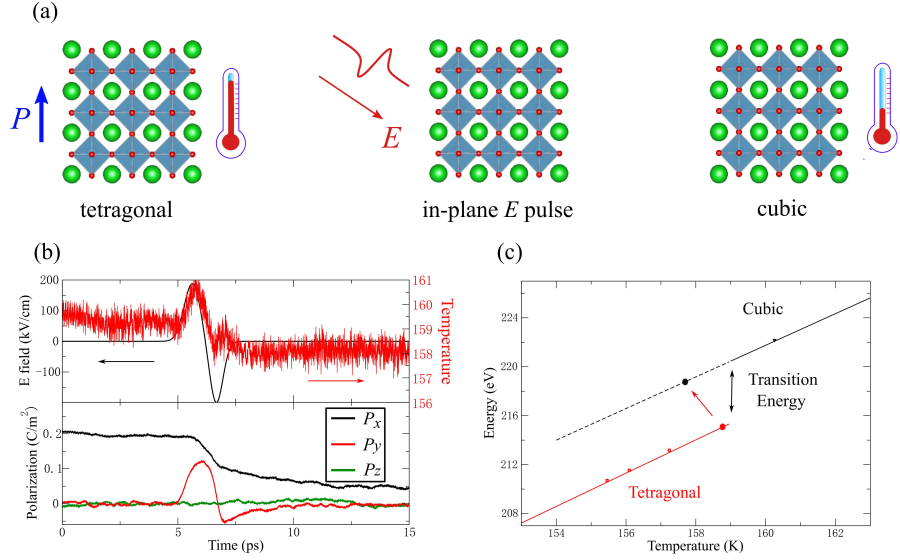} 
\caption{(a) Schematic representation of the computational experiment: using single--cycle THz electric field pulse perpendicular to the polarization in tetragonal BaTiO$_3$ gives a large reproducible negative ECE.
(b) Electric field pulse, temperature, and polarization along the three Cartesian axes.
(C) Energy {\em{vs.}} temperature of BaTiO$_3$. The red and black dots represent the structures before and after the electric field pulse respectively.
} \label{f_tc}
\end{figure}

\begin{figure}[htbp]
\centering\includegraphics[width=15.0cm]{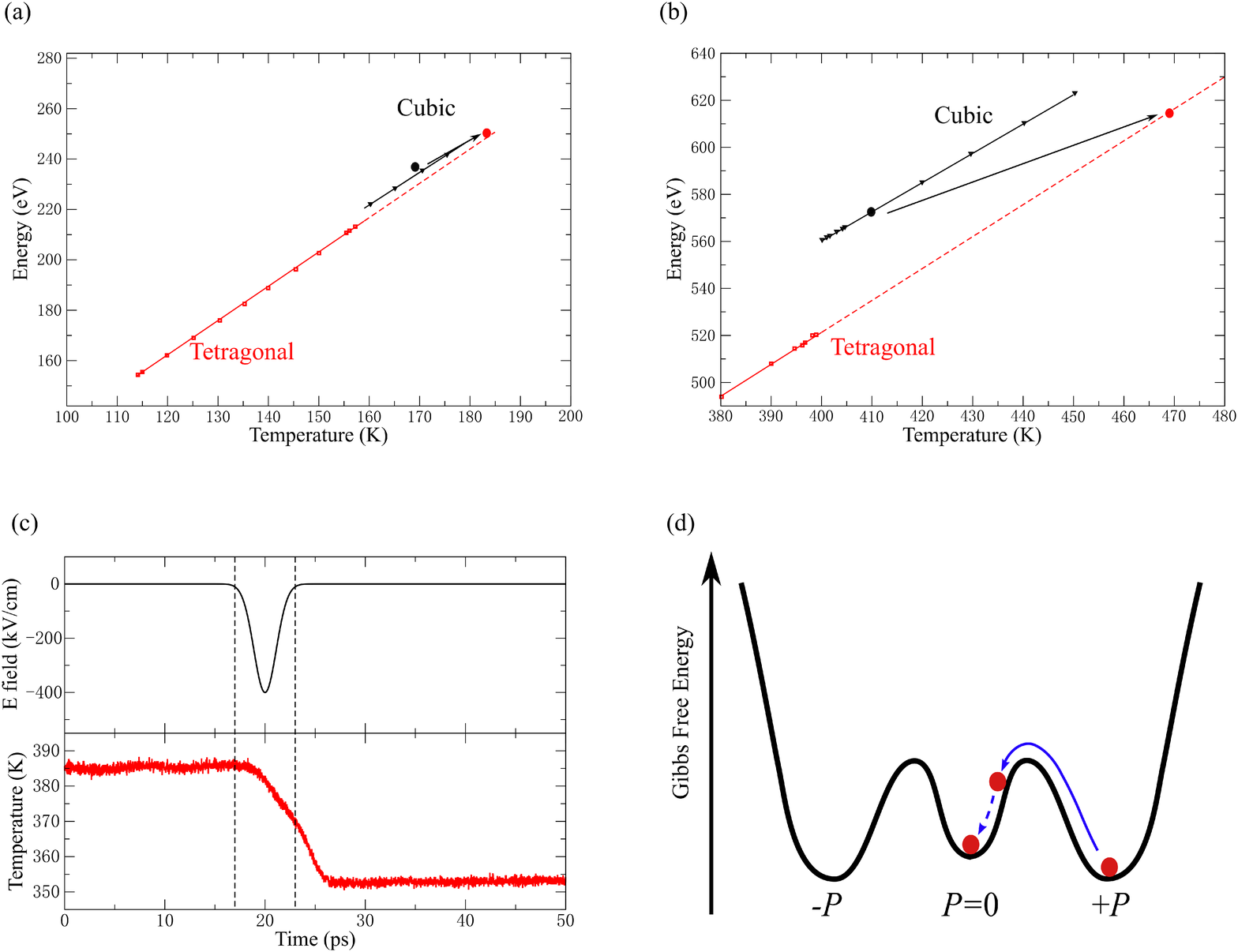} 
\caption{(a) Internal energy {\em{vs.}} temperature of BaTiO$_3$. A 600 kV/cm electric field induces a 12.7 K temperature increase.
(b) Internal energy {\em{vs.}} temperature of PbTiO$_3$. A 600 kV/cm electric field induces a 58.2 K temperature increases;
(c) A half--cycle THz electric field pulse anti--parallel to polarization in PbTiO$_3$ produces a giant (35 K) negative ECE;   
(d) Schematic representation of the polar phase evolution induced by the electric field pulse. 
The outer two minima of the free energy profile represent the crystal with positive and negative polarization. 
The central minimum represents a cubic phase. 
The solid line represents that the electric field pulse drives PbTiO$_3$ crystal from its tetragonal phase toward the cubic, and over the energy barrier.
The dashed line indicates that the system evolves to the cubic phase after the pulse. 
Due to the high energy barrier, the system can be trapped in the cubic local minimum for nanoseconds, which is long enough for thermal release.
} \label{f_bpto}
\end{figure}

\end{document}